\documentclass[ aip,jmp,amsmath,amssymb,reprint,]{revtex4-1}
\usepackage[utf8x]{inputenc}
\usepackage{graphicx}
\usepackage{epsfig}
\usepackage{amsmath}
\usepackage{amsfonts}
\usepackage{amssymb}
 \usepackage{afterpage}

\begin{document}

\preprint{AIP/123-QED}

\title[]{The two gap transitions in  Ge$_{1-x}$Sn$_x$: effect of non-substitutional complex defects }

\author{J. D. Querales-Flores}
\affiliation{Centro At\'omico Bariloche-CNEA and CONICET, Av. Bustillo Km. 9.5,  8400-Bariloche, Argentina}
\affiliation{Instituto Balseiro, Univ. Nac. de Cuyo and CNEA, 8400-Bariloche, Argentina}
\author{C. I. Ventura}
\affiliation{Centro At\'omico Bariloche-CNEA and CONICET, Av. Bustillo Km. 9.5, 8400-Bariloche, Argentina}
\affiliation{Univ. Nac. de R\'io Negro, 8400-Bariloche, Argentina}
\author{J. D. Fuhr}
\affiliation{Centro At\'omico Bariloche-CNEA and CONICET, Av. Bustillo Km. 9.5, 8400-Bariloche, Argentina}
\affiliation{Instituto Balseiro, Univ. Nac. de Cuyo and CNEA, 8400-Bariloche, Argentina}
\author{R. A. Barrio}
\affiliation{Instituto de F{\'{\i}}sica, U.N.A.M., Ap. Postal 20-364, 01000 M\'exico D.F., Mexico.}

\date{\today}

\begin{abstract}

% (236 words , here)

The existence of non-substitutional  $\beta$-Sn defects in Ge$_{1-x}$Sn$_{x}$ was confirmed by emission channeling 
experiments [Decoster \textit{et al.}, \textit{Phys. Rev. B} \textbf{81},  155204 (2010)], which established that although most Sn enters substitutionally ($\alpha$-Sn) in 
the Ge lattice, a second significant fraction corresponds to the Sn-vacancy defect complex in the split-vacancy configuration ( $\beta$-Sn ), 
in agreement with our previous theoretical
study [Ventura  \textit{et al.}, \textit{Phys. Rev. B} \textbf{79}, 155202 (2009)]. Here, we present our electronic structure calculation
for  Ge$_{1-x}$Sn$_{x}$, including substitutional $\alpha$-Sn as well as non-substitutional $\beta$-Sn defects. 
To include the presence of non-substitutional complex defects in the electronic
structure calculation for this multi-orbital alloy problem, we extended the  approach for the purely 
substitutional alloy by Jenkins and Dow 
 [Jenkins and Dow,  \textit{Phys. Rev. B} \textbf{36},  7994 (1987)]. We employed an effective substitutional two-site
 cluster equivalent to the real non-substitutional $\beta$-Sn defect, which was determined by a  Green's functions calculation. We then calculated 
 the electronic structure of the effective alloy purely in terms of substitutional defects,
  embedding the effective substitutional clusters  in the lattice. 
  Our results describe the two transitions of the fundamental gap of Ge$_{1-x}$Sn$_{x}$ as a function of the total Sn-concentration: 
  namely from an indirect to a direct gap, first, and the metallization transition
 at higher $x$. They also highlight the role of $\beta$-Sn in the reduction of the concentration range which corresponds
  to the direct-gap phase of this alloy, of interest for optoelectronics applications.

\end{abstract}


\maketitle

\section{Introduction}
The semiconductor technology based on Si has limitations for optoelectronic and photovoltaic device applications, 
related to the indirect nature of the fundamental bandgap  which results  in inefficient absorption and emission of light. To overcome these limitations,
direct energy-gap materials based on group IV semiconductors have been searched.\cite{Hull-1999, Menendez-2002, Menendez-2006, Goodman-1982}
Among group-IV elements, Ge is considered an important candidate to replace Si in semiconducting applications.\cite{Menendez-2002}
Compared to Si, Ge has a larger free-carrier mobility and a lower dopant activation temperature,\cite{Hull-1999}
which makes it an attractive material in future metal-oxide semiconductor field-effect transistors.\cite{Yeo-2005, Yang-2007}

Ge$_{1-x}$Sn$_{x}$  also  attracted considerable attention because it becomes a direct bandgap semiconductor
above $\sim$6-10\% Sn without external mechanical strain. The  tunability of its  gap with composition makes Ge$_{1-x}$Sn$_{x}$ 
a highly interesting material for infrared applications, especially at low Sn concentrations
($x<0.20$)\cite{Atwater}. Theoretical calculations indicated that strained Ge$_{1-x}$Sn$_{x}$  ($x<0.17$) would exhibit
enhanced electron and hole mobility, which could make the alloy also interesting for high-speed integrated circuits.\cite{Sau-2007,liu-APE2015}  
The integration of Ge in Si-based photonics is important for advances in the performance of detectors, modulators, and emitters.
  Recently there have been reports of room temperature direct bandgap emission for Si-substrate-based Ge  p-i-n  heterojunction photodiode structures,
   operated  under forward bias.\cite{kasper2011} 
   A temperature-dependent photoluminescence (PL) study has been conducted\cite{Du2014} in Ge$_{1-x}$Sn$_{x}$
films with Sn compositions of 0.9$\%$, 3.2$\%$, and 6.0$\%$ grown on Si. The competition between the direct and
indirect bandgap transitions was clearly observed. The relative peak intensity of the direct transition with
respect to the indirect transition increases with an increase in temperature, indicating the direct transition dominates the PL at high temperatures.
Furthermore, as Sn composition increases, a progressive enhancement of the PL intensity corresponding to the direct transition 
was observed,\cite{Du2014} due to the reduction of  the direct-indirect valley separation,
which experimentally confirms that  Ge$_{1-x}$Sn$_{x}$ grown on Si becomes a group IV-based direct bandgap material by increasing the Sn content. 
More recently, the fabrication and  properties of  Ge$_{1-x}$Sn$_{x}$ ($x<0.123$) $pn$ diodes (LEDs) were reported.\cite{menendez2015} 
Electroluminiscence results indicated that emission properties depend very sensitively on the Sn-concentrations on both sides of the junctions, 
making this system not only a serious candidate for laser devices but also an ideal model system to study the properties of quasi-direct light emitting devices. 

Experimental studies in group-IV alloys were hindered for a long time by sample preparation problems. 
When Ge$_{1-x}$Sn$_{x}$  samples are experimentally prepared, 
the distribution of the Sn atoms in the Ge matrix depends on the growth conditions:  Sn-atoms can enter randomly, 
form a regular superstructure, or coalesce into a larger cluster.  Below 13$^{\circ}$C, pure Sn exists in the $\alpha$-Sn (gray tin) phase with diamond structure, 
but it undergoes a phase transition to $\beta$-Sn (white-tin) above this temperature.\cite{cardona1985}
Experiments have shown that a problem with the incorporation of Sn  into the Ge lattice
 is the large $\sim$ 17$\%$ lattice mismatch between these elements, and the instability
of the diamond-cubic structure of $\alpha$-Sn above 13$^{\circ}$C.\cite{Menendez-2006,Gurdal} 
Although Ge$_{1-x}$Sn$_{x}$ alloys have been successfully grown by molecular beam epitaxy \cite{Lin-TSF2012},
chemical vapor deposition \cite{Menendez-2006}, 
and enhanced direct bandgap luminescence has been demonstrated for up to 8\% Sn \cite{Chen2011, Menendez2012},
at higher Sn-concentrations several limitations arise due to the low thermodynamic solid solubility of Sn
in the Ge crystal, which is less than 1$\%$ and in many cases the material quality is questionable due to the propensity of Sn 
to segregate toward the film surface.\cite{Menendez-2006,Gurdal,Wegscheider}
Recently, the fabrication of high quality Ge$_{1-x}$Sn$_{x}$ on InGaAs buffer layers 
using low-temperature growth by molecular beam epitaxy was reported.\cite{Lin-TSF2012} 
X-ray difraction, secondary ion mass spectroscopy and transmission electron microscopy studies, 
demostrated that up to 10.5$\%$ Sn had been incorporated into Ge$_{1-x}$Sn$_{x}$  thin films without
Sn precipitation.\cite{Lin-TSF2012}  
More recently, for Ge$_{1-x}$Sn$_{x}$ homogeneous epitaxial layers were grown on InP substrates for  the range $ 0.15 < x < 0.27 $.\cite{Nakatsuka2013}  
 the direct bandgap  for  the range $ 0.15 < x < 0.27 $ was obtained through photon absorption spectra 
measured with Fourier transform infrared spectroscopy. 

Experimentally, a series of values have been reported for the critical Sn concentration for the indirect to direct gap transition, 
hereafter denoted  $x_{c}^{I}$ : 
starting from  Atwater \textit{et al.} \cite{Atwater} whose optical absorption experiments predicted   $ 0.11 < x_{c}^{I} < 0.15 $, 
followed by Ladr\'on de Guevara \textit{et al.}\cite{Ladron2007} who reported  $ 0.10 < x_{c}^{I} < 0.13 $ from transmittance measurements,  
D'Costa  \textit{et al.}\cite{Menendez-2006} reported $x_{c}^{I} \sim 0.11 $ with  ellipsometry experiments,
and more recently Chen \textit{et al.}\cite{Chen2011} 
reported  $x_{c}^{I} \sim 0.07$ using photoluminescence. Then, 
 $x_{c}^{I} \sim  0.06 - 0.08 $ was suggested, based on photoluminescence studies of strain free-GeSn layers \cite{Chen2011, Menendez2012}.
 In 2013 Ryu  \textit{et al.}\cite{Ryu2013}, by means of temperature-dependent photoluminescence experiments
of Ge/Si and Ge$_{1-y}$Sn$_{y}$/Si, indicated  a possible indirect-to-direct bandgap transition at Sn content $\sim 0.06$, consistent with
DFT calculations in Ref. \onlinecite{Ying2008}.
More recently,  by a fit  to experimental data in Ge$_{1-x}$Sn$_{x}$  using  a theoretical model of the  bandgap bowing,
Gallagher  \textit{et al.}\cite{gallagher2014}  
estimated  a crossover concentration  of $x_{c}^{I}=0.09$,  significantly increased  from earlier estimations 
based on a strictly quadratic compositional dependence of the bandgaps in Ref.\onlinecite{jiang2014},
 who obtained $x_{c}^{I}=0.073$.

The incorporation of Sn in the Ge matrix has been investigated theoretically by Ventura \textit{et al.}\cite{CFB-2009, CFB-Physica} 
Through local defect electronic calculations, the formation of several complex Sn-defects in Ge$_{1-x}$Sn$_{x}$ alloy was analyzed: 
confirming that at low Sn concentrations substitutional $\alpha$-Sn,  in which a Sn atom occupies the position of a Ge atom
in the diamond lattice,  is favoured. Above a certain critical Sn concentration\cite{CFB-2009, CFB-Physica} dependent on temperature,  
Sn could also appear as the non-substitutional $\beta$-Sn complex defect, in which an interstitial Sn-atom occupies the center of a divacancy in
the Ge lattice.  Metallic Sn clusters resulting in inhomogeneous defect structures
could appear at still higher Sn concentrations.\cite{CFB-2009}  In 2010  emission channeling experiments
by S. Decoster \textit{et al.}\cite{decoster} confirmed the existence of $\beta$-Sn defects in the homogeneous Ge$_{1-x}$Sn$_{x}$ alloy,  
establishing that they represented the second significant fraction of Sn incorporated in the Ge lattice,  most Sn atoms entering substitutionally ($\alpha$-Sn) in Ge.
The existence of such a defect in amorphous Ge-Sn alloys had already
been confirmed by detailed M\"ossbauer experiments,\cite{Chambouleyron} which in fact showed a signal corresponding to a Sn atom in an octahedral environment,
besides the expected signal of the  tetrahedral environment  corresponding to substitutional $\alpha$-Sn. The local environment and the interactions of $\alpha$-Sn,
the Ge vacancy and $\beta$-Sn, confirming that $\beta$-Sn could be formed by natural diffusion of a vacancy around $\alpha$-Sn 
because of the small energy barrier for the process, have been also studied.\cite{ourJAP2013} 

In this work, we concentrate on the electronic structure calculation for Ge$_{1-x}$Sn$_{x}$, including substitutional $\alpha$-Sn as well as the 
non-substitutional $\beta$-Sn defects. At present, ab-initio electronic structure calculations can include individual non-substitutional defects, such as interstitials, 
but no standard approaches are available to tackle the problem posed by complex non-substitutional defects formed by many components. 
An example of such complex defects would be an interstitial impurity atom attached
to a divacancy, as is the case for ``$\beta$-Sn'' defects in Ge.  
In order to take into account the non-substitutional complex defects in the electronic structure calculation for this multi-orbital alloy problem, 
we transformed the real alloy problem into an equivalent purely substitutional effective alloy  problem. For this, 
we employed an effective substitutional two-site cluster  equivalent to the real non-substitutional
$\beta$-Sn defect, and  extended the approach originally proposed by Jenkins and Dow \cite{Jenkins} for the purely substitutional alloy, 
who used  20 tight-binding (TB) $sp^{3}s^{*}$ orbitals for the group IV elements
combined with the virtual crystal approximation (VCA) for substitutional disorder. 

This paper is organized as follows. We present  in Sec. \ref{metodo} a brief description of our proposal 
for  the inclusion of $\beta$-Sn non-substitutional defects in the electronic structure calculation. 
 In Sec. \ref{comparison} we discuss the indirect-to-direct gap transition in Ge$_{1-x}$Sn$_{x}$, as described by 
our  present extension of the TB+VCA approach, adjusted to experimental data. In Appendix A we show how the present TBA+VCA approach 
allows to improve the theoretical description of experimental  direct gap results  for substitutional  Ge$_{1-x-y}$Si$_{x}$Sn$_{y}$  ternary alloys.
 The results of our present electronic structure  calculation for binary Ge$_{1-x}$Sn$_{x}$  with the present approach,  
 including  non-substitutional $\beta$-Sn  as well as substitutional $\alpha$-Sn,  are presented in Sec. \ref{results}.
 In Sec. \ref{Conclusions} we summarize the conclusions of our work.

\section{Inclusion of $\beta$-Sn non-substitutional complex defects
in the electronic structure calculation}
\label{metodo}

Not existing any electronic structure calculation for Ge$_{1-x}$Sn$_{x}$  taking into account the complex non-substitutional $\beta$-Sn defects, which appear 
above a critical Sn concentration \cite{CFB-2009,decoster}, we devised an analytical  method to include them based on 
an extension of the virtual crystal approximation. A first implementation was presented in Ref.\onlinecite{slafes}, 
though the present proposed method includes improvements allowing us to refine our description of  Ge$_{1-x}$Sn$_{x}$.
 VCA assumes a random alloy to be composed of ``virtual'' atoms,
 forming a periodic crystal potential modelled as a composition-weighted average of the constituent element potentials.
  To include $\beta$-Sn 
in the electronic structure calculation, we extend the TB+VCA approach by Jenkins and Dow\cite{Jenkins} for the substitutional Ge$_{1-x}$Sn$_{x}$ alloy.
In the latter,  only substitutional $\alpha$-Sn  was assumed to be present, and a 20-orbital TB basis (s,p,s* states) was introduced for group IV elements.
The $20\times20$ Jenkins-Dow tight-binding Hamiltonian includes: second-neighbor\cite{Jenkins,Newman} and spin-orbit interactions.\cite{Jenkins,chadi77, CC-1973, CC-1975}

\begin{figure}
\centering
\includegraphics[width=8.6cm]{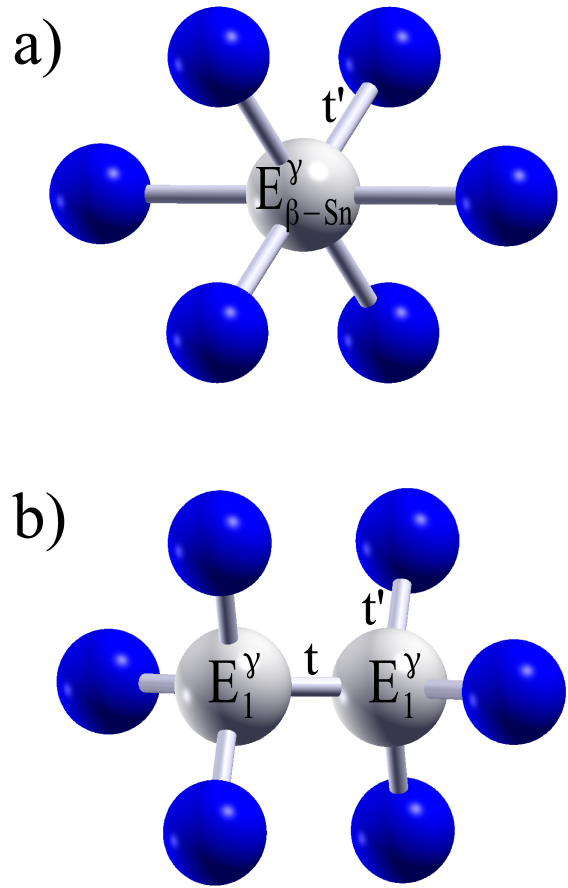}
\caption{\textbf{a)} Non-substitutional $\beta$-Sn, \textbf{b)} substitutional 2-site cluster equivalent for $\beta$-Sn.
Blue balls: Ge atoms (lattice with diamond symmetry); gray balls: Sn atoms, (a):Sn-atom located at the center of a Ge-divacancy and (b):Sn-atoms in substitutional positions. $t'$ represents a hopping between a Sn-atom and nearest neighbors Ge-atoms. $t$ denoted an intracluster hopping between Sn atoms on substitutional representation.}\label{esquema}
\end{figure}

For our present electronic structure calculation,
we  start by considering  the Ge$_{1-x}$Sn$_{x}$ alloy  formed by three components: Ge, $\alpha$-Sn and $\beta$-Sn atoms,
as an effective binary alloy composed by two components:  one, represented
by  the (Ge+$\alpha$-Sn) substitutional alloy as considered by Jenkins and Dow\cite{Jenkins},  and the other component
represents the $\beta$-Sn non-substitutional defects.   For the latter, we propose
an effective substitutional 2-site cluster equivalent to the real non-substitutional $\beta$-Sn\cite{slafes}, as illustrated in Fig. \ref{esquema} and 
 detailed in next subsection \ref{equivalencia}. In subsection II.B we present our extension of the Jenkins-Dow TB+VCA calculation to treat the effective substitutional 
 binary alloy representing the real alloy with Ge, $\alpha$-Sn and $\beta$-Sn.

\subsection{Effective substitutional two-site cluster equivalent to the non-substitutional $\beta$-Sn defect}\label{equivalencia}

As mentioned above, in a previous work\cite{slafes}  we determined  and compared two effective substitutional two-site clusters equivalent
 to the real non-substitutional $\beta$-Sn defects: schematically represented in  Fig.\ref{esquema}. 
 In Fig.\ref{esquema}.(a) we depict 
the real non-substitutional $\beta$-Sn defect ( with energy $E_{\beta-Sn}^{\gamma}$, 
where $\gamma$ represents each basis orbital ($\gamma = s, p, s^{*}$), 
in its sixfold coordinated configuration). Meanwhile    Fig.\ref{esquema}.(b) represents 
an equivalent cluster composed by two substitutional
sites, where the effective atoms occupying each site in
the cluster have an energy denoted by $E_{1}^{\gamma}$ (considered to be
equal in both sites of the cluster, by symmetry).\cite{slafes} 

The equivalence was established under the following conditions: 
(1) for simplicity, we propose that the equivalence is valid 
for each separate orbital; 
(2) we assume that only interactions between orbitals of the same type 
between nearest-neighbour (NN) atoms are relevant; 
and (3) we demand that the local Green's functions in the original and equivalent problems 
are equal, and will thus have the same analytical properties.
 
 Due to our investigation in Ref.\onlinecite{slafes},  
no qualitative differences are expected for the electronic structure obtained 
using equivalent clusters  with null intra-cluster hopping $ t = 0 $, or with 
  intra-cluster hopping equal to that between the cluster and the rest of
the lattice: $ t = t'$.

 Thus, in our present work we adopt the effective equivalent substitutional cluster with $ t = 0 $,   
to represent the $\beta$-Sn non-substitutional defects. In this case, from 
Ref. \onlinecite{slafes} we have:

\begin{equation}\label{solt1}
 E_{1}^{\gamma}\simeq  (E_{\beta-Sn}^{\gamma} + t' )
\end{equation}

The energy of the effective pseudo-atoms with the correct symmetry has been expressed in terms of energy
parameters corresponding to the original $\beta$-Sn defect, allowing to incorporate
these non-substitutional defects in the electronic structure calculation of the
alloy under study in terms of substitutional defects. 

\subsection{TB+VCA extension to include $\alpha$-Sn and non-substitutional  $\beta$-Sn in the Ge$_{1-x}$Sn$_{x}$ electronic structure calculation}
\label{our-VCA}

To obtain the virtual-crystal band structure of Ge$_{1-x}$Sn$_{x}$, including substitutional  $\alpha$-Sn  and non-substitutional $\beta$-Sn, 
 we propose the following  extension for 
the TB+VCA approximation of the substitutional alloy  originally proposed by Jenkins and Dow.\cite{Jenkins} 
We will use the same basis of 20 tight-binding orbitals  ( with $s$, $p$, $s^{*}$ character)  introduced by Jenkins and Dow for the group IV elements, 
but  in the diagonal part of the Hamiltonian   we will add a set of  three  orbital weight factors,  depending on the orbital character of the basis state, 
and take into account the presence of $\beta$-Sn.  Concretely, we consider the following matrix elements for the TB+VCA Hamiltonian 
of the binary alloy:

\begin{eqnarray}
\label{newTB}
\mathcal{H}_{ii}^{\gamma}  & =  & (1-x) \, [Ge]_{ii}^{\gamma} \, + \,  W_{\gamma}\left(x_{\alpha} \, [\alpha-Sn]_{ii}^{\gamma}  + \, x_{\beta} \, [\tilde{\beta}-Sn]_{ii}^{\gamma} \right); 
\nonumber \\                                                  
\mathcal{H}_{ij}^{\gamma} &  =  & \frac{ \left[ (1 - x ) [Ge]_{ij}^{\gamma} \{ a_{Ge} \}^{2} + x [\alpha-Sn]_{ij}^{\gamma} \{ a_{Sn} \}^{2} \right]}{ \{ a(x) \}^{2}   }         
\raisetag{2pt}
\end{eqnarray}

where $\mathcal{H}_{ii}^{\gamma}$ and $\mathcal{H}_{ij}^{\gamma}$ respectively, denote the diagonal and non-diagonal matrix elements of the Hamiltonian, 
subindices $i$ and $j$ refer to the TB-orbital states and $\gamma$ denotes each Hamiltonian block with $s$, $p$ or $s^{*}$
orbital character. Further details of the structure of this TB Hamiltonian  can be found in Ref.\onlinecite{Jenkins}.
By [Ge]  and [$\alpha$-Sn] we  respectively refer to the tight-binding parameters  for  pure Ge  or  for  $\alpha$-Sn,  as given in Ref. \onlinecite{Jenkins}. 
Meanwhile  [$\tilde{\beta}$-Sn]  denotes the TB Hamiltonian matrix elements corresponding to the substitutional equivalent used for the real non-substitutional defect, 
in our case:  [$\tilde{\beta}$-Sn] = $E_{1}^{\gamma}$ = [$\beta$-Sn] + $t'$.   In our present calculations, for simplicity we have assumed 
that a Sn-atom has the same tight-binding parameters in both configurations: $\alpha$-Sn and $\beta$-Sn. 
By $x_{\alpha}$  and $x_{\beta}$  we denote the relative concentration of $\alpha$-Sn and $\beta$-Sn respectively in Ge$_{1-x}$Sn$_{x}$,
therefore: $ x = x_{\alpha} + x_{\beta}$. 

In the diagonal matrix elements  of the Hamiltonian above, 
we introduced three  orbital weight factors: $W_{\gamma}$ ($\gamma$ = $s$,$s^{*}$ or $p$),  in order to reproduce as closely as posible the band structure 
of Ge$_{1-x}$Sn$_{x}$ alloys,  according to  recent experiments,\cite{Atwater, Menendez-2006, gallagher2014}
and in particular  improve the description of the indirect to direct gap transition  of the substitutional alloy w.r. to the original approach by Jenkins and Dow \cite{Jenkins}.
$W_{\gamma}$   were included as  factors of the Sn-atoms contribution to the diagonal matrix elements, 
as the pure Ge indirect (and direct) gaps\cite{handbook} are correctly described by the Jenkins and Dow TB elements\cite{Jenkins}.
In  following subsection we will discuss in detail  the  parametrization  adopted for  these  $W_{\gamma}$ parameters.

Notice that we adopt the non-diagonal Hamiltonian matrix elements of Jenkins and Dow,\cite{Jenkins} which include the lattice parameters for Ge 
and  $\alpha$-Sn, namely: $a_{Ge}=5.65$ \AA,  and $a_{Sn}=6.46$ \AA, and we are assuming 
that Vegard's law\cite{vegard} is valid for the binary alloy  lattice parameter, thus: $ a(x) = (1-x) a_{Ge} + x a_{Sn} $.

The present extension of  TB+VCA  enables us to tackle two important issues in Ge$_{1-x}$Sn$_{x}$: 
1) as already mentioned,  a more realistic description of the crossover from indirect to direct fundamental bandgap, 
according to recent experiments,  and 
2) the inclusion of non-substitutional complex $\beta$-Sn defects in the electronic structure calculation, which have been confirmed to exist 
by experiments, and as we will show would play an important role in the electronic properties of Ge$_{1-x}$Sn$_{x}$, 
basically limiting  the direct gap phase  of interest for optoelectronics applications.

\subsubsection{Parametrization of W$_{\gamma}$:  indirect to direct gap  transition in substitutional Ge$_{1-x}$Sn$_{x}$}
\label{comparison}

Now, we will explain how we optimized the values of the weight parameters:  $W_{\gamma}$ ($\gamma$ = $s$,$s^{*}$ or $p$), 
introduced in Eq.\ref{newTB}, in order  to properly reproduce the available experimental data of Ge$_{1-x}$Sn$_{x}$ alloys, 
and in particular  the critical Sn-concentration  $x_{c}^{I}$ for  the indirect to direct gap  transition in  substitutional Ge$_{1-x}$Sn$_{x}$.

Using the original TB+VCA approach for substitutional Ge$_{1-x}$Sn$_{x}$ by Jenkins and Dow\cite{Jenkins}
 one obtains that  $x_{c_{-JD}}^{I}$ = 0.15\cite{CFB-2009}.   Recently, other theoretical predictions were reported:
  $x_{c}^{I}$= 0.17  was obtained with a charge self-consistent pseudo-potential plane wave 
 method\cite{Moontragoon2007},   $x_{c}^{I}$= 0.11 with the empirical pseudopotential method with adjustable form factors
   fitted to experimental data\cite{Low2012}, while the full potential augmented 
plane wave plus local orbital method within density functional theory(DFT) yielded:    $x_{c}^{I}$= 0.105.\cite{Chibane2010}
Gupta \textit{et al.}\cite{pseudo2013} predicted  $x_{c}^{I}$  =$0.065$, using a theoretical model based on the nonlocal empirical pseudopotential
method.  Eckhardt  \textit{ \textit{et al.}},\cite{Eckhardt2014} 
predicted  $x_{c}^{I} \sim$  $0.1$ for Ge$_{1-x}$Sn$_{x}$ grown commensurately on Ge(100) substrates, 
using a supercell approach and VCA, with DFT in the local density approximation.

Our  extension of  the TB+VCA described in subsection \ref{our-VCA} includes three orbital weight parameters
$W_{\gamma}$ ($\gamma$ = $s$, $s^{*}$ or $p$), which we introduced to improve the fit of the TB model 
to experimental electronic structure results. They  enable us to construct a TB model to 
correctly reproduce the experimental data reported  for the indirect and
direct bandgaps of Ge and Ge$_{1-x}$Sn$_{x}$  at particular alloy concentrations.\cite{Atwater,Menendez-2006,Chen2011}
 In our present  work,  we adjusted  $W_{\gamma}$ ($\gamma$ = $s$,$s^{*}$ or $p$),
using experimental data for the  indirect and direct bandgaps of pure Ge ($x=0$) and Ge$_{0.85}$Sn$_{0.15}$: obtained 
by He and Atwater \cite{Atwater} through optical absorption experiments, 
and confirmed nine years later by transmittance measurements by D'Costa \textit{et al.}\cite{Menendez-2006}
The set of optimal  weight factors obtained by fitting these data are:
\begin{equation}
 W_{s} = 1.256  \quad , \quad W_{s^{*}} = 1.020 \quad ,  \quad  W_{p}=1.00 
\label{weights}
\end{equation}

Regarding the choice of the weight factors specified in Equation \ref{weights}, it was done taking into account 
our study of the orbital character of  the eigenvectors along the Brillouin zone (BZ),\cite{slafes}
 and in particular at the symmetry points 
which define the fundamental bandgap, depending on the alloy composition: namely the BZ center, $\Gamma$, 
with the relevant valence band maximum, and the conduction band minimum determining the direct gap, and BZ point $L= ( 2\pi/ a(x) ) (1/2, 1/2, 1/2)$
where the conduction band  minimum determining the indirect gap in Ge lies. 
 At $\Gamma$, we find that the eigenvectors correspond mainly to $s$-states at the conduction band minimum 
 and mainly $p$-states at the valence band maximum, while at $L$ the eigenvectors correspond mostly to $(s+s^{*})$-states, though a minor proportion of $p$-character is retained, 
 and these results do not exhibit significant changes increasing Sn-concentration. Meanwhile, increasing $x$   the energy of the valence band maximum at $\Gamma$
  remains almost fixed, while it is the energies of the conduction band minima at $\L$ and $\Gamma$ which are changed.\cite{slafes} 
  Based on these facts we set $W_{p}=1.00$, and proceeded to adjust $W_{s}$ and $W_{s^{*}}$  
  to fit the experimental direct and indirect bandgaps\cite{Atwater,Menendez-2006,Chen2011} at $x={0.15}$.

Henceforward, the set of values for $W_{\gamma}$ specified in Eq.\ref{weights}   will be kept constant  for all  electronic structure calculations in this work. 
Notice that the only change to the matrix elements of the tight-binding Hamiltonian of Jenkins and Dow\cite{Jenkins}, 
due to the introduction of these optimized weight factors $W_{\gamma}$ in Eq.\ref{newTB}, 
appears in the diagonal TB parameters for $\alpha$-Sn (and $\beta$-Sn in our approximation), here given by:  
$E_{s} = -7.3853$eV and $E_{s*} = 6.0180$eV.

% In Table \ref{factores}, we  show the rest of the TB parameters used in our calculation to construct the $20\times20$ TB+VCA Hamiltonian $\mathcal{H}$
% as in Ref.\onlinecite{Jenkins}.
% Notice that the adjusted\cite{Atwater, Menendez-2006}  weight factors $W_{\gamma}$ ($\gamma=s,s^{*},p$) 
% slightly  modified the values of the diagonal TB parameters $E_{s}$, and $E_{s*}$
% corresponding to $\alpha$-Sn and $\beta$-Sn in Table \ref{factores}, 
% w.r. to those in Ref. \onlinecite{Jenkins}.

%\begin{table}[h!]
%\centering
%\caption{$\alpha$-Sn  and Ge tight-binding parameters (in eV) except $d$, which is in \AA). The notation is that of Jenkins and Dow (Ref. \onlinecite{Jenkins}).}
%\begin{tabular*}{8 cm}{c @{\extracolsep{\fill}}  c@{\extracolsep{\fill}}c}
%\begin{tabular}{10 cm}{c    |    c c    |      c   }
%\hline \hline 
%       &  Ge  & $\alpha$-Sn, $\beta$-Sn (This work)   \\
% \hline 
% $E_{s}$ &  -5.8800 &  -7.3853  \\

%$E_{p}$ & 1.5533  &  1.1733  \\

%$\lambda$ & 0.0967 & 0.2667 \\

%$E_{s*}$ &  6.3900 &  6.0180    \\

%$U_{ss}$ & -6.7800 &  -5.4600 \\

%$U_{xx}$ & 1.6500 &  1.4400   \\

%$U_{xy}$ & 4.8416  &  3.9042  \\

%$U_{sp}$ & 4.9520 &  4.0172  \\

%$U_{s*p}$ &  4.5030 & 3.6459   \\

%$W_{ps'}$ &  0.1352  &  0.1229   \\

%$d$ &  2.45  &  2.81   \\

%\hline \hline

%\end{tabular*}

%\label{factores}
%\end{table}

In Table \ref{bandgaps}, we show a comparison between the direct and indirect bandgaps obtained with the TB+VCA approach described above 
and the experimental results reported in Ref. \onlinecite{Atwater} for them.
As expected,  there is agreement for $x=0$ and $x=0.15$, used for our fits, but at intermediate Sn concentrations, 
the calculated values show some deviations from the experimental ones, as discussed in next section. 

\begin{table}[h!]
\centering
\caption{Comparison between the theoretical bandgap energies in the present work, and 
the reported experimental data of Atwater et al. \cite{Atwater} for Ge$_{1-x}$Sn$_{x}$ . $E_{0}$: Direct gap  and  $E_{I}$: Indirect gap. }
\begin{tabular}{c | c | c | c| c}
\hline \hline
$x$ & $E_{0}:$ Exp. & $E_{0}:$ Theory & $E_{I}:$ Exp. & $E_{I}:$ Theory \\
\hline \hline
0.00 & 0.800 $\pm$ 0.004 &  0.803 & 0.670$\pm$0.019  &  0.670 \\

0.06 &  0.614 $\pm$ 0.004   &  0.613 & 0.599$\pm$0.019  &  0.578 \\

0.11 &  0.445 $\pm$ 0.003  &  0.468 & 0.428 $\pm$ 0.019  &  0.502\\

0.15 & 0.346 $\pm$ 0.003  & 0.346 & 0.441 $\pm$ 0.004  &  0.441 \\
\hline \hline

\end{tabular}

\label{bandgaps}
\end{table}

A considerable improvement is obtained for the predicted critical Sn-concentration  for the transition
 from  an indirect to  a direct fundamental gap  in Ge$_{1-x}$Sn$_{x}$, 
which the present approach  places in the vecinity of $ x_{c}^{I} = 0.088$ ( with the optimized  $W_{\gamma}$ values, specified in Eq.\ref{weights}).   
This is much closer  to the recently reported experimental values detailed in the Introduction
 than the prediction  of $x_{c_{-JD}}^{I} \sim 0.15$,  obtained\cite{CFB-2009} using the original TB+VCA by Jenkins and Dow\cite{Jenkins}.

 %  REVISAR ESTO con las refs. 
 Finally, before discussing in detail the results obtained with our approach in next section, 
 we would like to comment on two recent parametrizations of the tight binding parameters,\cite{Kufner2013,Attaoui2014} 
 and compare them with those including spin orbit interaction and up to second-nearest neighbour effects proposed by Jenkins and Dow in 1987.\cite{Jenkins} 

  In 2013, K\"{u}fner \textit{et al.},\cite{Kufner2013} studied the structural and electronic properties 
 of pure  $\alpha$-Sn  nanocrystals from first principles, using DFT within approximations based on the hybrid exchange-correlation functional
and including spin-orbit interaction effects. They reported a  list of first-NN  tight-binding parameters for $\alpha$-Sn 
with some differences  with respect to those of Refs.\onlinecite{Vogl,Jenkins}.  
Replacing in the TB+VCA approach by Jenkins and Dow\cite{Jenkins}  the first-NN tight-binding parameters of pure $\alpha$-Sn by those of K\"{u}fner \textit{et al.}\cite{Kufner2013},  
we find that the prediction for the indirect to direct gap transition in Ge$_{1-x}$Sn$_{x}$ would be shifted from $x_{c_{-JD}}^{I} \sim 0.15$\cite{CFB-2009, Jenkins} to $x_{c}^{I} \sim 0.10$. 
 
 In 2014, based on the first-NN  tight-binding parameters for pure Ge of Vogl et al.\cite{Vogl} and those by K\"{u}fner \textit{et al.} for $\alpha$-Sn\cite{Kufner2013}, 
  Attiaoui et al.\cite{Attaoui2014} presented a semi-empirical second-NN tight-binding calculation of
   the electronic structure of Ge$_{1-x-y}$Si$_{x}$Sn$_{y}$ ternary alloys as well as Ge$_{1-x}$Sn$_{x}$.
   Using the same 20 basis states as Refs.\onlinecite{Vogl,Jenkins,Kufner2013}, they evaluated   
new TB parameters  for the alloy components, with some changes to the previous ones also including second-neighbours and spin orbit corrections\cite{Jenkins}. 
For the indirect to direct gap transition in substitutional unstrained Ge$_{1-x}$Sn$_{x}$ they obtain 
 $x_{c}^{I}\sim 0.11$, which is smaller than $x_{c_{-JD}}^{I} \sim 0.15$ and 
agrees with the experimental data of Ref. \onlinecite{Menendez-2006},  
though it overestimates more recent experimental values mentioned in the Introduction.\cite{Chen2011,Menendez2012,gallagher2014} 
Concerning the electronic structure study of substitutional unstrained Ge$_{1-x-y}$Si$_{x}$Sn$_{y}$ ternary alloys, 
the prediction of Ref.\onlinecite{Attaoui2014} is consistent with previous results\cite{ternarios} 
obtained using an extension of the TB+VCA by Jenkins and Dow\cite{Jenkins}. In particular, almost equal critical concentration values
 (with minor differences possibly related to the use of different pure Sn-lattice parameter values\cite{ternarios,Attaoui2014})
are predicted for the transition of the indirect gap: between  two different relevant conduction band minima  defining the gap (one Ge-like at  Brillouin zone point L, and the other Si-like).

\section{Results and discussion}\label{results}

Here, we will  present electronic structure results obtained using our approach described in previous section, divided in two subsections. 
In the first one, we will discuss and compare in detail with  recent experiments our results for the low Sn-concentration regime: 
 where only substitutional $\alpha$-Sn is expected to be present.  In the second subsection,  we show our predictions for  Ge$_{1-x}$Sn$_{x}$ alloys
  at higher Sn-concentrations where,  above a temperature-dependent critical Sn-concentration, 
   also non-substitutional $\beta$-Sn defects are expected to be present\cite{CFB-2009,CFB-Physica}. 
   In particular, for each of these two regimes, 
   we discuss the  fundamental gap transition of  Ge$_{1-x}$Sn$_{x}$  which is expected to take place. 
  
\subsection{Substitutional alloy: results of the present approach for Ge+$\alpha$-Sn } 
\label{subs}

\begin{figure}[!h]
  \begin{center}
  \includegraphics[width=8.6cm]{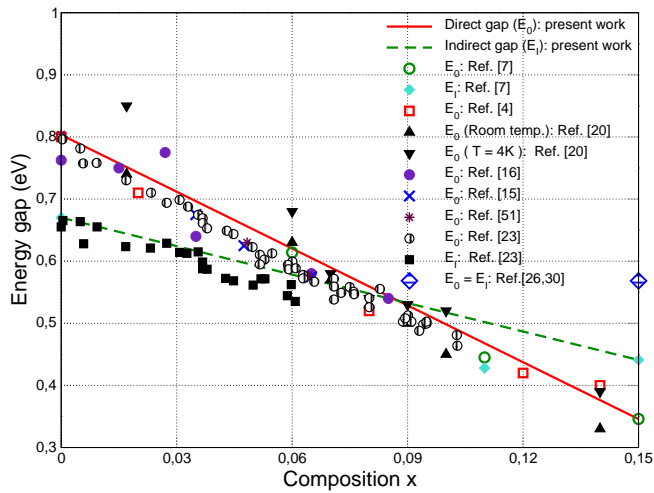}
    \caption[]{ (Color online) Comparison between our theoretical bandgap energies ( direct gap: full line, indirect gap: dashed line) yielding:  $ x_{c}^{I} = 0.088$, 
     and available experimental data ( symbols: see refs. in the plot), for various Sn compositions.  The theoretical gap 
     prediction at critical $ x_{c_{-JD}}^{I} = 0.15 $ obtained\cite{CFB-2009} with the original TB+VCA  approach \cite{Jenkins} is also shown (large diamond).
    %  Refs. \onlinecite{Menendez-2006,Atwater,Lin-TSF2012,Chen2011,gallagher2014,Chen-APL2012}. 
    }
  \label{ajustes}
  \end{center}
  \end{figure}

Using the TB+VCA approach  described in previous section for Ge$_{1-x}$Sn$_{x}$ alloys in the low $x$ regime,   
where all Sn atoms occupy substitutional positions in the Ge lattice, we obtained 
the results exhibited in Fig. \ref{ajustes}.  
There,  we show the compositional  dependence of the direct ($E_{0}$) and indirect ($E_{I}$)  bandgaps  of Ge$_{1-x}$Sn$_{x}$ obtained with the present approach, 
intersecting at $ x_{c}^{I} = 0.088$ where the indirect to direct  transition of the fundamental gap is thus predicted.
For comparison, we included a series of  recently available experimental data,\cite{Menendez-2006,Atwater,Lin-TSF2012,Chen2011,Ladron2007,gallagher2014,Chen-APL2012}
and also show the indirect to direct gap transition prediction of $ x_{c_{-JD}}^{I} = 0.15$, obtained using the original TB+VCA approach,\cite{Jenkins,CFB-2009}
which yields a gap value of 0.568 eV at that Sn-concentration, much higher than the observed  fundamental direct gap of 0.346 eV\cite{Atwater}
 detailed in Table \ref{bandgaps}.

The  direct and indirect  bandgaps obtained with our approach (in eV) can be respectively  fitted as follows:
 
\begin{eqnarray}  
E_{0} (x)  & = & 0.803 - 3.047 \,  x   \quad  ;                                                    
\nonumber  \\    
E_{I} (x)  & = & 0.670 - 1.527  \, x    \quad .                        
\label{fits}
\end{eqnarray}

 Comparing our results with the experimental data in  Fig. \ref{ajustes},   one sees that a relatively good description of the  
  compositional dependence of the direct and indirect gaps is obtained, 
  indeed much better than if a linear interpolation between the respective gap values for 
  pure Ge and $\alpha$-Sn was used.\cite{Atwater, Menendez-2006, Chen2011}
  Nevertheless, small deviations from linearity in the data are evident,\cite{Atwater,Menendez-2006,Menendez-2011}  as experimentally reported 
  also for the lattice constant  of Ge$_{1-x}$Sn$_{x}$\cite{Menendez-2011}. 
  Many other binary semiconducting alloys A$_{x}$B$_{1-x}$ also exhibit similar non-linear dependences of their physical properties as a function of alloy composition, 
behaviour known as \textit{bowing effects}. Though VCA  cannot describe non-linear bowing effects, it nevertheless often yields good qualitative results.\cite{Menendez-2006}

% parameters in Table \ref{factores} and the adjusted $W_{\gamma}$ weight factors
%, we found that increasing the Sn-content in Ge$_{1-x}$Sn$_{x}$ 
%both the indirect and direct bandgaps are reduced, 

% and the electronic structure shows the following features: 

% ii) Secondly, at higher Sn-content, the binary alloy undergoes  a transition from a direct gap semiconductor to a metallic regime.
% This metallization transition will be the main subject of subsection \ref{metal}. 

%First, we check the validity of our extended TB+VCA approach by comparing the band structure and total density of states (DOS) obtained 
%for substitutional Ge$_{1-x}$Sn$_{x}$, with those obtained using the \textit{ab-initio} FPLO5+CPA code\cite{Koepernik} in Ref.\onlinecite{CFB-2009}.

  \begin{figure}[h!]
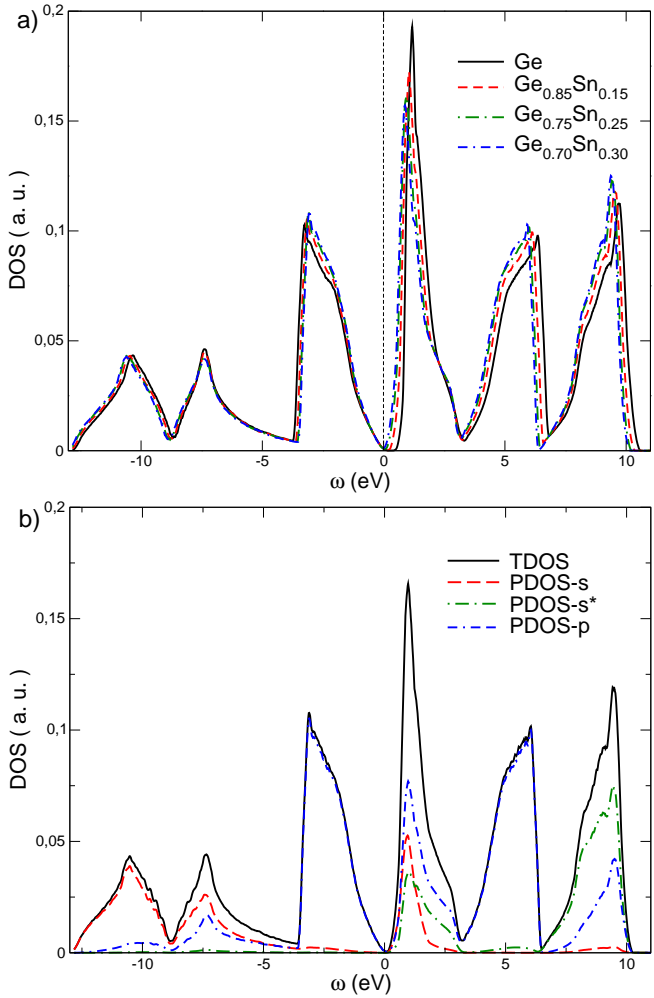

  \begin{center}
  \includegraphics[width=8.6cm]{Figure3a.eps}
  \includegraphics[width=8.6cm]{Figure3b.eps}
  \caption[]{(Color online). Substitutional  Ge$_{1-x}$Sn$_{x}$ - results of the present TB+VCA:  
  a) Total density of states for the  $\alpha$-Sn concentrations indicated in the plot; b) Total (TDOS) and partial (PDOS) ``$p$'',``$s$ and $s^{*}$'' 
  densities of states at $x=0.22$ ( $ >  x_{c}^{I} = 0.088$),  a direct gap alloy. $t'= -3$eV as in Refs.\onlinecite{Barrio-1986, Barrio-RMF}.}
  \label{proyVCA}
  \end{center}
  \end{figure}

%\begin{widetext}

Fig.\ref{proyVCA}(a)  shows the total density of states  obtained for Ge$_{1-x}$Sn$_{x}$ as a function of energy, at  substitutional-Sn concentrations:
$x= 0.0, 0.15, 0.25, 0.30$.  In agreement with the results obtained using the \textit{ab-initio} FPLO5+CPA code\cite{Koepernik} in Ref.\onlinecite{CFB-2009},  
notice that a smooth behaviour as function of Sn concentration is obtained, with changes in the bandwidth, and a progressive
reduction of the gap around the $\omega = 0$ with Sn concentration.  Experiments in Ge$_{1-x}$Sn$_{x}$\cite{Atwater, Menendez-2006, Ladron2007, Chen2011, Chen-APL2012} 
confirmed that the direct gap decreases primarily through a Sn-content increase in the alloy.\cite{Atwater, Menendez-2002, Menendez-2006}

In Fig. \ref{proyVCA}(b), we show the total  and the three partial densities of states  as a function of energy for Ge$_{0.78}$Sn$_{0.22}$ substitutional alloy.
Around the bandgap two peaks are visible in the density of states:  the peak  located just below the gap is clearly dominated by $p$ orbital contributions,
while   $p, s$ and $s*$ orbitals contribute to the peak located  just above the gap.
Moreover, we can see that the lowest band ($\sim$ -12 eV) corresponds essentially to $s$ orbitals, while the highest band ($\sim$ 10 eV) is originated 
essentially by $s*$ and $p$ orbitals, as in pure Ge.  These features also agree with the FPLO5+CPA results\cite{CFB-2009, Koepernik}.

%\begin{figure}[h!]
%\vspace{1cm}
%  \begin{center}
%  \includegraphics[width=8.6cm]{Figures/Fig4-v2.eps}
%    \caption[]{(Color online) Substitutional TB+VCA for Ge$_{1-x}$Sn$_{x}$: total density of states for the $\alpha$-Sn concentrations indicated in the plot.
%    Parameters used for the present  TB+VCA extension: as in Fig. \ref{proyVCA}. }
%  \label{TDOS}
%  \end{center}
%  \end{figure}

%\end{widetext}

  \begin{figure}[h!]
  \begin{center}
  \includegraphics[width=8.6cm]{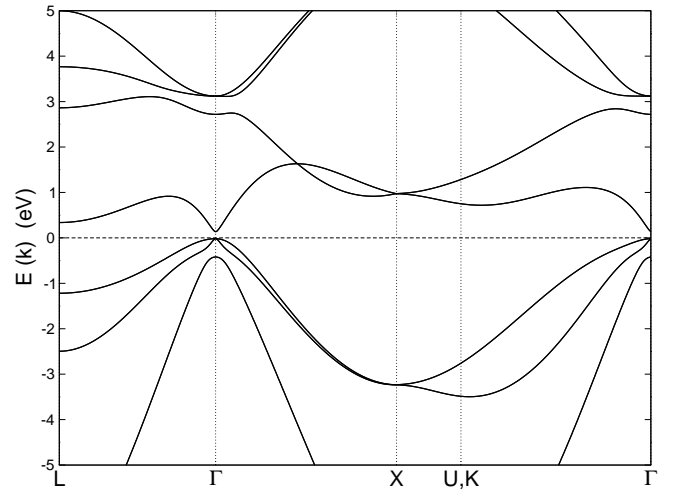}
  \caption[]{(Color online) Ge$_{0.78}$Sn$_{0.22}$ substitutional alloy: band structure $E(\vec{k})$ obtained along 
  BZ paths of the face-centered cubic diamond lattice, between symmetry points: $\Gamma = (0, 0, 0)$, $L = (2\pi/a)(1/2, 1/2, 1/2)$,
  $X = (2\pi/a)(1, 0, 0)$, $U = (2\pi/a)(1, 1/4, 1/4)$, and $K = (2 /a)(3/4, 3/4, 0)$, being $a$ the lattice parameter. Parameters used for
   the present  TB+VCA: as in Fig. \ref{proyVCA}.}
  \label{banda-sust}
  \end{center}
  \end{figure}

 In Fig. \ref{banda-sust}, we show the band structure obtained for the substitutional  Ge$_{0.78}$Sn$_{0.22}$ alloy. 
 In order to analyze the fundamental bandgap in Ge$_{1-x}$Sn$_{x}$,  we focus on three specific BZ points which define it: 
 the maximum of the valence band at $\Gamma$,
 the minima of the conduction band at $\Gamma$ and at $L$,  with energies denoted $\Gamma_{0}$, $\Gamma_{1}$ and $L_{1}$ respectively.
 Notice that Ge$_{0.78}$Sn$_{0.22}$ possesses a fundamental direct gap :  0.132 eV  $ = \Gamma_{1} - \Gamma_{0}$ , 
 and as indirect gap:  0.334 eV $ = L_{1} - \Gamma_{0}$.  Moreover,  analyzing the orbital-character  of the band structure: we found that the eigenvectors 
corresponding to $\Gamma_{0}$ and $\Gamma_{1}$, are mainly due to $p$ and $s$  states, respectively. Meanwhile, the eigenvectors at $L_{1}$
 are due to $s$ and $s^{*}$ states. These results  are analogous to those obtained for pure Ge, which indicates that the relative weights 
  of the orbital contributions to the electronic properties of Ge$_{1-x}$Sn$_{x}$  are weakly dependent of Sn-content  in the  substitutional alloy.

\subsection{Ge$_{1-x}$Sn$_{x}$: TB+VCA results including $\alpha$-Sn and $\beta$-Sn, metallization transition}\label{metal}

In previous section, we showed that  Ge$_{1-x}$Sn$_{x}$ possesses an indirect fundamental gap at low Sn-concentrations, and that 
 by increasing Sn-concentration the  substitutional binary alloy  undergoes a crossover from an indirect to a direct bandgap at $x_{c}^{I}\sim0.088$. 
At higher Sn-concentrations,   non-substitutional  $\beta$-Sn defects appear in the binary alloy, as predicted theoretically\cite{CFB-Physica, CFB-2009} 
and confirmed by experiments \cite{decoster}.  We will now show that the presence of $\beta$-Sn  reduces  the concentration range 
where Ge$_{1-x}$Sn$_{x}$ possesses a direct gap,  i.e. it reduces the critical concentration $x_{c}^{II}$
at which the direct gap closes,  corresponding to the metallization transition.

 We analyzed the effect of both  substitutional and non-substitutional Sn defects  on the electronic structure of Ge$_{1-x}$Sn$_{x}$ 
  employing the TB+VCA extension proposed in the present work, detailed in Section \ref{metodo}.

\begin{figure}[h!]
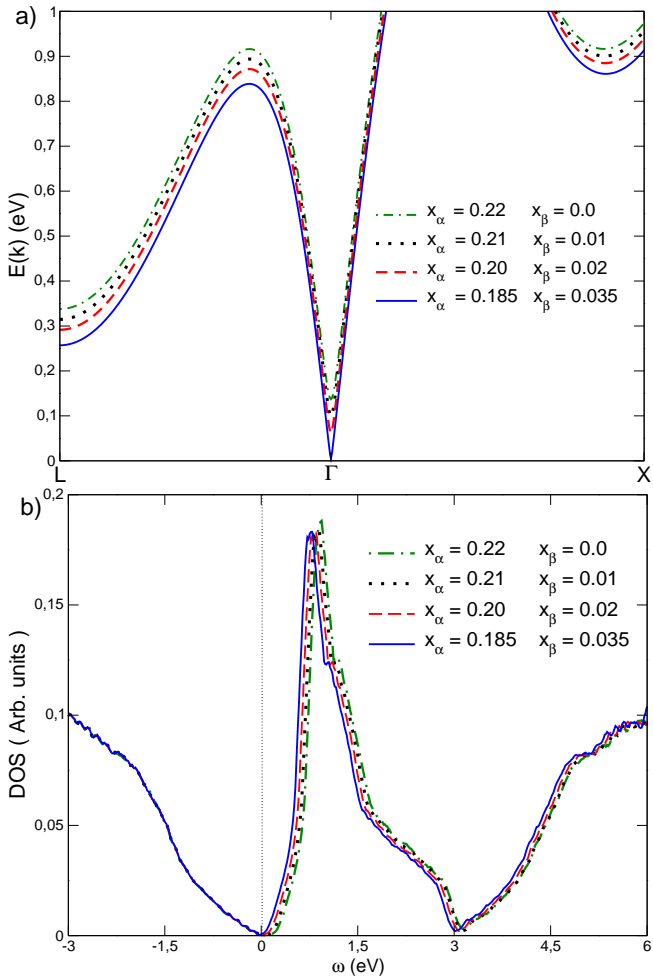

\begin{center} 
\includegraphics[width=8.6cm]{Figure5a.eps}
\includegraphics[width=8.6cm]{Figure5b.eps}
\caption[]{(Color online)  a) $\beta$-Sn effect on the band structure of Ge$_{0.78}$Sn$_{0.22}$: the lowest conduction band
along $L-\Gamma-X$ path for different values $x_{\beta}$  of $\beta$-Sn content, as indicated in the plot.  b) effect of $\beta$-Sn
on the DOS of Ge$_{0.78}$Sn$_{0.22}$: relative $\beta$-Sn similar to Fig. 6(a).  
Parameters used for  the present  TB+VCA extension: as in Fig. \ref{proyVCA}.
}\label{bandasSn22}.
\end{center}
\end{figure}

To assess the effect of $\beta$-Sn defects on the electronic structure, in particular,  on the bandgaps,
in Fig. \ref{bandasSn22}(a), we plot the TB+VCA band structure of  Ge$_{0.78}$Sn$_{0.22}$ along the 
$L-\Gamma-X$ Brillouin zone path, for  a fixed total Sn-concentration  $x= x_{\alpha} + x_{\beta} = 0.22$, 
but  different relative  contents of $\alpha$-Sn and  non-substitutional $\beta$-Sn.  
Notice that when $x_{\beta}$  increases, for example from 0 to 0.035,  the bandgap decreases:    a progressive 
reduction of the direct gap at $\Gamma$ upon increase of the  $\beta$-Sn concentration  is observed.  In Fig. \ref{bandasSn22}(b),
we show the effect of $\beta$-Sn upon  the DOS,  for fixed  total Sn-concentration $x=0.22$ and varying the  relative $\beta$-Sn content in the alloy. 
The changes  induced  by  increasing  $\beta$-Sn concentration on the density of states  are less noticeable 
 than  those  obtained  in  the  band structure (see Fig. \ref{bandasSn22}(a)).
The  progressive reduction of the gap confirms  that  the presence of non-substitutional $\beta$-Sn favours the metallization in Ge$_{1-x}$Sn$_{x}$.

\begin{figure}[h!]
\begin{center} 
\includegraphics[width=8.6cm]{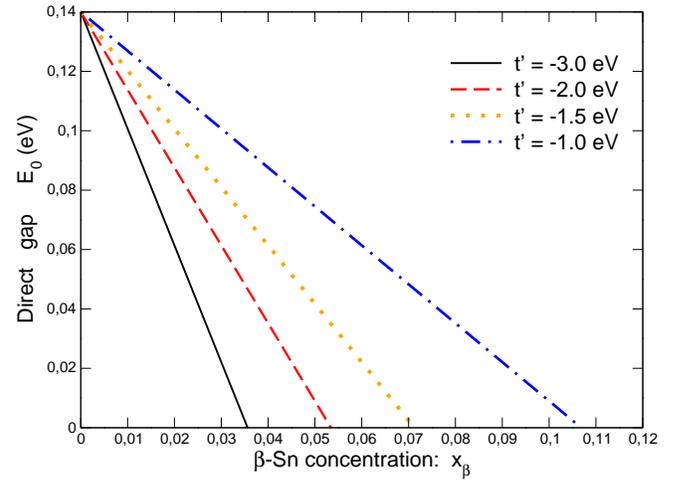}
\caption[]{(Color online) $t'$ dependence of the metallization transition for Ge$_{0.78}$Sn$_{0.22}$: 
energy value at $\Gamma_{0}$ ($E_{0}$) as a function of $x_{\beta}$, for different values of $t'$.
Other parameters used for  the present  TB+VCA extension: as in Fig. \ref{proyVCA}
}\label{Sn22}
\end{center}
\end{figure}

In Fig. \ref{Sn22}  we plot the direct gap energy $E_{0}$,  for the binary alloy with fixed total Sn-concentration $x = 0.22 $
as a function of  $x_{\beta}$  ( i.e. $ x_{\alpha} = 0.22 - x_{\beta}$ ),  for four different values of hopping $t'$ between $\beta-$Sn and its Ge nearest-neighbours. 
A linear dependence  of $E_{0}$  as a function of $x_{\beta}$  is obtained. The critical concentration $x_{c}^{II}$ at which the direct gap closes, 
and the alloy becomes metallic,  is strongly dependent on $t'$  in our approach. 

Analyzing  the effect of  non-substitutional  $\beta$-Sn  on the total and partial densities of states, using our TB+VCA approach 
 we  find that  the relative weights of the orbital contributions to the electronic properties of Ge$_{1-x}$Sn$_{x}$  
 are weakly dependent of Sn-content.

%\begin{figure}[h!]

% \vspace{0.7cm}
%  \begin{center}

%  \includegraphics[width=8.6cm]{Figures/Fig8-v2.eps}
%    \caption[]{(Color online). Full Ge$_{1-x}$Sn$_{x}$ calculation (including $\beta$-Sn): orbital dependence of TB+VCA density of states. 
%Total DOS and partial ``p'',``s'' and ``s*''densities of states at: x=0.22 and relative defect contents: $x_{\alpha}$=0.19 and $x_{\beta}$=0.03. 
%Parameters used for the present  TB+VCA extension: as in Fig. \ref{proyVCA}. 
%}
%  \label{DOS-betas}
%  \end{center}
%  \end{figure}

 Experimentally, the critical Sn-concentration   $x_{c}^{II}$ for the metallization transition  in  Ge$_{1-x}$Sn$_{x}$  is yet unknown.
 As shown in Fig.\ref{ajustes}  gap measurements were reported at relatively low Sn-concentrations:  
$x < 0.15 $\cite{Atwater, Menendez-2006, Low2012, Chen2011, Lin2011, Ladron2007}. More recently, the direct gap was determined 
in   homogeneous epitaxial  layers  of Ge$_{1-x}$Sn$_{x}$  grown on  InP substrates \cite{Nakatsuka2013}  
 for   $ 0.15 < x < 0.27 $, from photon absorption spectra measured with Fourier transform infrared spectroscopy. 
 In particular, for  Ge$_{0.73}$Sn$_{0.27}$ Nakatsuka et al.\cite{Nakatsuka2013} obtained a direct gap of $\sim$0.25 eV, 
so that  the metallization of these epitaxial layers  would occur at  $ x > 0.27 $.

Additional measurements  for  higher Sn-concentrations  would be required  to locate  the metallization transition in bulk Ge$_{1-x}$Sn$_{x}$. 
Extrapolating experimental data fits, 
two predictions for  $x_{c}^{II}$ were obtained:  $x_{c}^{II}\sim0.30$ by  Atwater \textit{ \textit{et al.}}\cite{Atwater},
and   $x_{c}^{II}\sim 0.37$ by V.R. D'Costa \textit{ \textit{et al.}}\cite{Menendez-2006}   

On the other hand, Jenkins and Dow predicted $x_{c}^{II}\sim0.62$ \cite{Jenkins}  with their original TB+VCA approach for substitutional  Ge$_{1-x}$Sn$_{x}$. 
If we  would consider  only substitutional $\alpha$-Sn   to be present and  use  our present TB+VCA extension,
we would predict  for the transition from a direct gap to a metallic regime: $x_{c}^{II} \sim 0.26$, 
same value which can be obtained if using the TB parameters for $\alpha$-Sn reported in Ref. \onlinecite{Kufner2013}. 

In 2009 we proposed a statistical model  for the formation of $\beta$-Sn  defects in Ge$_{1-x}$Sn$_{x}$, \cite{CFB-Physica,CFB-2009} 
from which the relative concentrations of $\beta$-Sn ( and  $\alpha$-Sn)  in the alloy
can be obtained  as  a function of temperature and  the total Sn-concentration $x$.

\begin{figure}[h!]
\begin{center} 
    \includegraphics[width=8.6cm]{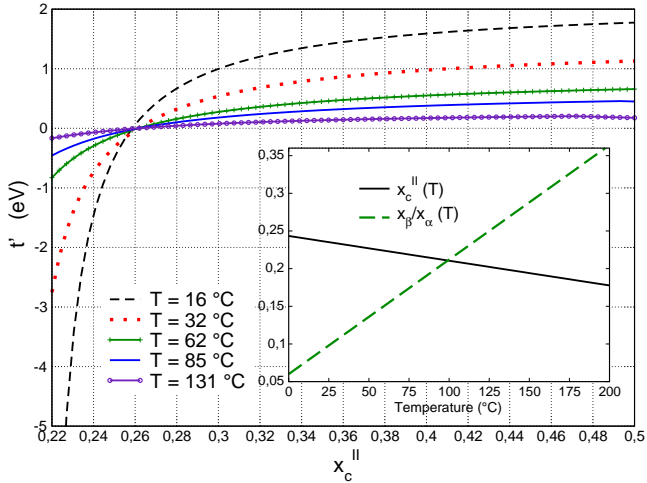}
    \caption{(Color online) $t'$ hopping values  as a function of the critical Sn concentration for metallization: $x_{c}^{II}$, at different temperatures.
    Inset: temperature dependence of the relative  proportion of non-substitutional to substitutional Sn, $x_{\beta}/x_{\alpha}$, at  $x= x_{c}^{II}$ 
    as obtained from the statistical model of Refs.\onlinecite{CFB-Physica, CFB-2009}; and, as a result, 
     our prediction for the temperature dependence of the metallization transition:  $x_{c}^{II} (T) $, for $t'= -3$eV.\label{transicion-dir-metal}}
\end{center}   
  \end{figure}

Using the statistical model \cite{CFB-Physica,CFB-2009}  to determine  $x_{\beta}$  in Ge$_{1-x}$Sn$_{x}$ ( and from it: $ x_{\alpha} = x - x_{\beta}$ ) , 
we can explore the  dependence of the critical concentration for metallization  $x_{c}^{II}$ 
on the only free parameter ($t'$)   in our present approach.
Furthermore,   if $x_{c}^{II}$  were experimentally known,   it might be used to  tune  the free parameter  $t'$. 
In Figure \ref{transicion-dir-metal} we  address this issue, by plotting $t'$ corresponding  to $x_{c}^{II}$ values in the 
range from $0.22$ to $0.5$ for different temperatures $T$. Specifically  for T = 16 $^{\circ}$C, 
32 $^{\circ}$C, 62 $^{\circ}$C, 85 $^{\circ}$C and 131 $^{\circ}$C,  while the 
relative contents $x_{\beta}$ and $x_{\alpha}$  are obtained from the statistical model\cite{CFB-Physica,CFB-2009}

%To demonstrate  the influence of $t'$ on $x_{c}^{II}$, we examine $t'$ in the $x_{c}^{II}$ range from $0.22$ to $0.50$ for different temperatures ($T$),

Figure \ref{transicion-dir-metal} shows  that for  $0.22 \leq x_{c}^{II} < 0.26 $,  $t'$  is negative  and 
 increases monotonically with increasing $x_{c}^{II}$. On the other hand, 
 for $x_{c}^{II}\geq 0.26$, $t'$ is positive. 
 Note that, all of these graphs in Fig. \ref{transicion-dir-metal} intersect at  $x_{c}^{II} = 0.26$, which corresponds to $t'=0$: 
 which in our model is equivalent to considering only substitutional Sn to be present.

For instance, if we consider the temperature $ T = 16^{\circ}$C, by adjusting $t'\sim 1.0037$ eV we would obtain a value $x_{c}^{II}=0.30$,  
as suggested extrapolating the experiments of Ref. \onlinecite{Atwater},
while, if we consider the same temperature but adjust  $t'\sim 1.4920$eV, we would obtain a value $x_{c}^{II}=0.37$, as suggested by extrapolating the 
experiments of Ref. \onlinecite{Menendez-2006}.

The inset of Figure \ref{transicion-dir-metal} depicts, for  $t' = - 3eV$,  the temperature dependence obtained in our approach 
 for the critical concentration for metallization, $x_{c}^{II}$ (full line),
 and for  the ratio between the non-substitutional $\beta$-Sn and the substitutional  $\alpha$-Sn (dashed line) concentrations at $x= x_{c}^{II}$ 
 from Refs.\onlinecite{CFB-Physica, CFB-2009}. 
 The inset reveals an increase in $x_{\beta}/x_{\alpha}$ as a function of temperature,
while a decrease in $x_{c}^{II}$ as a function of temperature is observed. 
These results indicate that increasing the non-substitutional Sn content leads to a lower $x_{c}^{II}$,  i.e. confirming that 
the presence of $\beta$-Sn favours metallization. 

As we mentioned in the Introduction, the main interest for technological applications of Ge$_{1-x}$Sn$_{x}$ is linked to its direct gap phase.\cite{Menendez-2002,Menendez-2006}
However, our results show that the concentration range ($\Delta x =  x_{c}^{II} - x_{c}^{I} $), in which Ge$_{1-x}$Sn$_{x}$ possesses 
a direct fundamental gap would be reduced if the temperature of formation
of the alloy is increased, since $\beta$-Sn might appear at lower Sn-concentrations.  An important aspect in the fabrication of high-quality thin films
using molecular beam epitaxy  is the growth-temperature\cite{Menendez-2002, Menendez-2006, Menendez-APL2011, Menendez-APL2010}.
Therefore, a detailed experimental study of the electronic properties of these alloys, including the second gap transition (metallization), 
could allow to determine  the optimal growth conditions  to  control  the proportion  of   $\beta$-Sn  in Ge$_{1-x}$Sn$_{x}$.

\section{Conclusions}\label{Conclusions}

We have studied  the effect of $\beta$-Sn non-substitutional complex defects
on the electronic structure of the Ge$_{1-x}$Sn$_{x}$ binary alloy. 

In order to include non-substitutional complex defects in an electronic structure calculation,
we presented our extension of the method 
originally proposed by Jenkins and Dow for substitutional Ge$_{1-x}$Sn$_{x}$,
using 20 tight-binding  $sp^{3}s^{*}$ orbitals for the group IV elements combined 
with the virtual crystal approximation.  We included the complex non-substitutional
$\beta$-Sn defects through the introduction of an effective  substitutional two-site equivalent
cluster for them, which we appropriately embedded in the lattice to calculate 
 the electronic structure of the effective alloy purely in terms of substitutional defects.

Our method allows to describe the two transitions of the fundamental gap of Ge$_{1-x}$Sn$_{x}$ as a function
of the total Sn concentration. In particular: i) with our proposed extension for the tight-binding matrix, 
we could tune the critical Sn concentration $x_{c}^{I}$ for the transition from an indirect to a direct gap,
 in agreement with the most recent experimental data. Our extension also improves the
  theoretical  description of the direct gap 
 experimental values for ternary Ge$_{1-x-y}$Si$_{x}$Sn$_{y}$ alloys, as shown in the Appendix.
ii) The metallization transition ( closure of the direct gap) of Ge$_{1-x}$Sn$_{x}$ at higher $ x = x_{c}^{II}$ 
can also be described, and we demonstrated the relevance
of non-substitutional  $\beta$-Sn for the determination of $x_{c}^{II}$. In fact, if $x_{c}^{II}$ would be measured 
it could be used to determine  the hopping parameter ($t'$) between $\beta$-Sn defects and the Ge matrix, 
only free parameter of our model.
iii) We predict the effect of temperature on the metallization transition $x_{c}^{II} (T)$ (not yet experimentally measured),
resulting from the increase with temperature of the ratio $x_{\beta}/x_{\alpha}$ between the concentrations of non-substitutional
to substitutional Sn in Ge (which we obtain from our statistical model for  Ge$_{1-x}$Sn$_{x}$).

We believe that the physical properties of Ge$_{1-x}$Sn$_{x}$ are strongly affected by
Sn complex defects in the Ge matrix , such as $\beta$-Sn among others\cite{CFB-2009}, and future experimental work 
would provide further support and increase the usefulness of the present approach.  Furthermore, the general idea 
we proposed to include non-substitutional complex defects in electronic structure calculations, 
by determining effective substitutional equivalent clusters  to replace them  
and solve the problem in terms of purely substitutional effective alloys, 
has a wide potential for  applications in other systems.

\section{Acknowledgments}

C.I.V. and J.D.F. are Investigadores Cient{\'{\i}}ficos of CONICET (Argentina). 
J.D.Q.F. has a fellowship from CONICET. C.I.V. acknowledges support for this work from CONICET (PIP 0702) and 
 ANPCyT (PICT'38357; PICT Redes'1776), and R.A.B. from CONACyT project 179616.

 \appendix
 
\section{Application to Ge$_{1-x-y}$Si$_{x}$Sn$_{y}$ ternary alloys}

 In 2013 we reported a first calculation of the electronic structure  of Ge$_{1–x–y}$Si$_{x}$Sn$_{y}$ ternary alloys\cite{ternarios}, 
 employing a combined TB+VCA approximation method  which we developed 
as a direct extension to ternary substitutional alloys of the Jenkins and Dow approach \cite{Jenkins} for binary substitutional Ge$_{1-x}$Sn$_{x}$.
The same problem was later studied  also in Ref.\onlinecite{Attaoui2014}.

\begin{figure}[h!]
\begin{center} 
    \includegraphics[width=8.6cm]{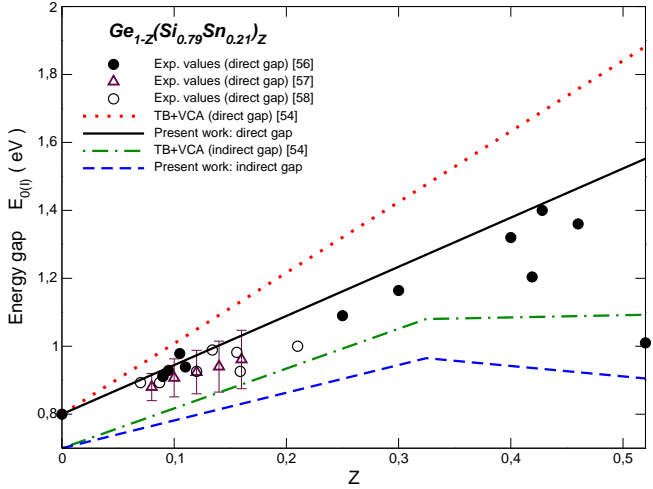}
    \caption{(Color online) Compositional dependence of the bandgap of 
Ge$_{1-Z}$(Si$_{0.79}$Sn$_{0.21}$)$_{Z}$ ternary alloys lattice-matched to Ge,  obtained using the modified TB+VCA approach in present work: direct gap (solid line) and indirect gap (dashed line). For comparison we included our results from Ref.\onlinecite{ternarios}: direct gap ( dotted line) and indirect gap (dash-dotted line). Other data included for comparison correspond to experimental results for the direct gap from Refs. \onlinecite{menendez-prl2009,beeler-apl2011,xu-jacs2012}.}
    \label{figure10}
\end{center}   
  \end{figure}

  \begin{figure}[h!]
\begin{center} 
    \includegraphics[width=8.6cm]{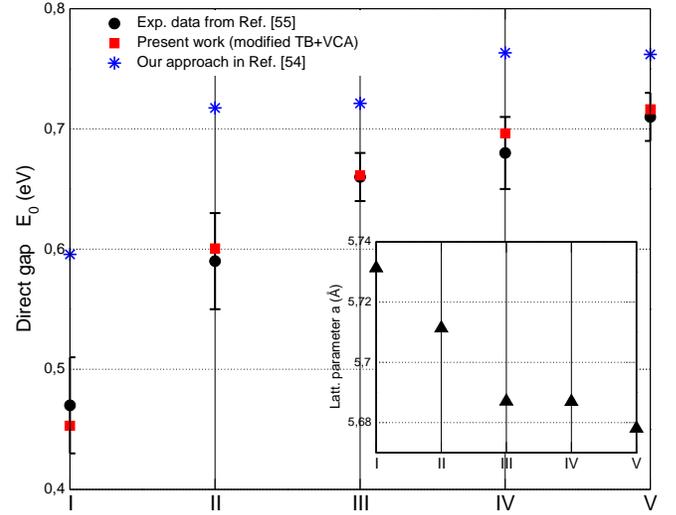}
    \caption{(Color online) Comparison of our theoretical results for the direct gap and lattice parameter with experimental data reported in Ref. \onlinecite{chi2013}, for the following ternary alloys: I: Ge$_{0.864}$Sn$_{0.10}$Si$_{0.036}$, II: Ge$_{0.86}$Sn$_{0.082}$Si$_{0.058}$, III:  Ge$_{0.94}$Sn$_{0.042}$Si$_{0.018}$, IV:  Ge$_{0.916}$Sn$_{0.047}$Si$_{0.037}$ and V:  Ge$_{0.947}$Sn$_{0.032}$Si$_{0.021}$. Inset: lattice parameter for each of the ternary alloys I-V.}
       \label{figure11}
\end{center}   
  \end{figure}

Our  electronic structure results  for ternary Ge$_{1–x–y}$Si$_{x}$Sn$_{y}$  confirmed 
predictions and experimental indications that a 1 eV bandgap was attainable with these alloys, 
as required for the fourth layer planned to be added to present-day record-efficiency triple-junction solar cells, 
in order to further increase their efficiency for satellite applications.

It is interesting to apply the present TB+VCA parametrization for the substitutional alloy discussed in Section  \ref{comparison} of the present work, 
in particular the renormalized $\alpha-$Sn tight-binding parameters, as starting point 
to recalculate  the electronic structure of the ternary alloys and compare it with the results of our previous calculation \cite{ternarios}. 
We exhibit  the comparison of the direct gap  results as a function of composition in Figs. \ref{figure10}  and  \ref{figure11} . 
In Fig. \ref{figure10}  one sees how  the parametrized  TB+VCA approach of the present work improves the description of the experimental 
compositional dependence of the direct gap w.r. to our previous approach of Ref.\onlinecite{ternarios} for a series of samples  lattice-matched to Ge, 
while we also show how the predicted  indirect gap is slightly modified.
In Fig. \ref{figure11} we compare our present parametrized  TB+VCA  results for the direct gap  with recent  experimental results\cite{chi2013} 
for five  ternary alloys with different lattice parameters ( shown as inset): within the experimental error bars  the agreement  is indeed 
very good for all samples, greatly improved w.r. to the direct gap values predicted by the previous TB+VCA approach\cite{ternarios}.

\end{document}